\documentclass[epj]{svjour}

\usepackage[dvips]{graphicx}
\usepackage{array}
\usepackage{amssymb}
\usepackage{amsmath}
\usepackage{hhline}
\usepackage{longtable}
\usepackage{dcolumn}
\usepackage{bm}
\usepackage{subfigure}
\usepackage{epsfig}
\usepackage{latexsym,amsmath}
\usepackage{amsbsy}
\begin{document}
\title{Topologies and Laplacian spectra of a deterministic uniform recursive tree}

\author{Zhongzhi Zhang\inst{1,2} \thanks{e-mail: zhangzz@fudan.edu.cn} \and Shuigeng Zhou\inst{1,2} \thanks{e-mail: sgzhou@fudan.edu.cn} \and Yi Qi\inst{1,2}  \and Jihong Guan\inst{3}}                     
\institute{Department of Computer Science and Engineering, Fudan
University, Shanghai 200433, China \and Shanghai Key Lab of
Intelligent Information Processing, Fudan University, Shanghai
200433, China \and Department of Computer Science and Technology,
Tongji University, 4800 Cao'an Road, Shanghai 201804, China}

\date{Received: date / Revised version: date}

\abstract{The uniform recursive tree (URT) is one of the most
important models and has been successfully applied to many fields.
Here we study exactly the topological characteristics and spectral
properties of the Laplacian matrix of a deterministic uniform
recursive tree, which is a deterministic version of URT. Firstly,
from the perspective of complex networks, we determine the main
structural characteristics of the deterministic tree. The obtained
vigorous results show that the network has an exponential degree
distribution, small average path length, power-law distribution of
node betweenness, and positive degree-degree correlations. Then we
determine the complete Laplacian spectra (eigenvalues) and their
corresponding eigenvectors of the considered graph. Interestingly,
all the Laplacian eigenvalues are distinct.
} 

 \maketitle
\section{Introduction}

The flexibility and generality in the description of natural and
social systems have made complex networks become an area of
tremendous recent interest~\cite{AlBa02,DoMe02,Ne03,BoLaMoChHw06}.
Among many interesting aspects, topological characterization is very
significant for the study in network field. In the past ten years,
there has been a considerable interest in characterizing and
understanding the topological properties of networked
systems~\cite{CoRoTrVi07}. A lot of network measurements have been
proposed, among which degree distribution, average path length
(APL), betweenness, and degree correlations have been extensively
studied, since they have profound effects on the dynamical processes
taking place on networks, such as
robustness~\cite{AlJeBa00,CaNeStWa00,CoErAvHa01,ZhZhZo07}, epidemic
spreading~\cite{PaVe01a,MoPaVe02},
synchronization~\cite{BaPe02,ZhRoZh06,CoGa07}, and
games~\cite{SzFa07,RoLiWa07}.

The above mentioned topological characteristics focus on direct
measurements of structural properties of networks. Apart from these
investigations there exists a vast literature related to (Laplacian)
spectrum of complex
networks~\cite{FaDeBaVi01,GoKaKi01,DoGoMeSa03,ChLuVu03,YaZhWa06},
which provides useful insight into the relevant structural
properties of graphs. In fact, topological features capture the
static structural properties of complex networks, while spectrum
provides global measures of the network
properties~\cite{FaDeBaVi01}. In the past years, graph spectrum has
found many important applications in physics and other
fields~\cite{Ne03,BoLaMoChHw06}. For example, the ratio of the
maximum eigenvalue to the smallest nonzero one of Laplacian matrix
determines the synchronizability of the
network~\cite{BaPe02,ZhRoZh06,CoGa07}. On the other hand, the
eigenvectors of Laplacian matrix have also been successfully used to
detect community structure of networks~\cite{Ne06}. While the
Laplacian eigenvalues and eigenvectors have high influence on the
structural properties of networks and dynamics running on them,
until now, most analysis of Laplacian spectra and eigenvectors has
been confined to approximate or numerical methods, the latter of
which is prohibitively time and memory consuming for large
networks~\cite{FaDeBaVi01}.

On the other hand, in order to describe real systems and study their
structural properties, a wide variety of network models have been
presented~\cite{AlBa02,DoMe02,Ne03,BoLaMoChHw06}, among which the
uniform recursive tree (URT) is perhaps one of the most widely
studied models~\cite{SmMa95}. It is now established that the URT is
one of the two principal models~\cite{DoKrMeSa08,ZhZhZhGu08} of a
random graph (the second one is the famous Erd\"os-R\'enyi
model~\cite{ErRe60}). The URT is perhaps the simplest tree and is
built in the following way: at each time step, we attach each new
node to an existing node which is chosen uniformly at random. It has
found applications in several areas. For example, it has been
suggested as models for the spread of epidemics~\cite{Mo74}, the
family trees of preserved copies of ancient or medieval
texts~\cite{NaHe82}, chain letter and pyramid schemes~\cite{Ga77},
to name but a few. Recently, a deterministic version~\cite{JuKiKa02}
of the URT has been proposed to mimic real-life systems whose number
of nodes increases exponentially with time. This kind of
deterministic models have drawn much attention from the scientific
communities and have turned out to be a useful tool
~\cite{BaRaVi01,DoGoMe02,CoFeRa04,ZhRoZh07,RaBa03,NaUeKaAk05,AnHeAnSi05,ZhRoCo06,CoOzPe00,ZhRoGo06,Hi07,BaCoDa06,ZhZhFaGuZh07,ZhZhZoChGu07,BeMaRo08}.
Although uniform recursive tree is well
understood~\cite{SmMa95,DoKrMeSa08,ZhZhZhGu08,Mo74,NaHe82,GoOhJeKaKi02,DoMeOl06},
less is known about the topologies and other nature of the
deterministic uniform recursive tree (DURT)~\cite{JuKiKa02}.

In this paper, we offer a detailed analysis of the deterministic
uniform recursive tree (DURT)~\cite{JuKiKa02} from the viewpoint of
complex networks.  We first determine accurately relevant
topological characteristics of DURT, such as degree distribution,
average path length, betweenness distribution, and degree
correlations. We then use methods of graph theory and algebra to
calculate or estimate the eigenvalues and eigenvectors of the
Laplacian matrix. We present that there is a strong relationship
between the eigenvalues and the eigenvectors of the Laplacian
matrix.

\section{The deterministic uniform recursive tree}

\begin{figure*}
\begin{center}
\includegraphics[width=18cm]{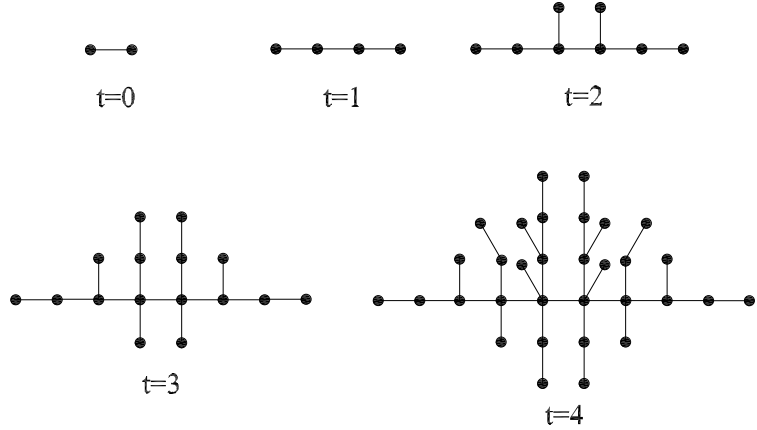}
\caption{Illustration of the deterministic uniform recursive tree,
showing the first five steps of growth process.} \label{net01}
\end{center}
\end{figure*}

The deterministic uniform recursive tree is one of the simplest
models. It is constructed in an iterative way~\cite{JuKiKa02}. We
denote the tree (network) after $t$ steps by $U_{t}$ ($t\geq 0$).
Then the network at step $t$ is built as follows. For $t=0$, $U_{0}$
is an edge connecting two nodes. For $t\geq 1$, $U_{t}$ is obtained
from $U_{t-1}$. We attach a new node to each node in $U_{t-1}$. This
iterative process is repeated, then we obtain a deterministic tree
with an exponential decreasing spectrum of degrees as shown below.
The definition of the model is illustrated schematically in
Figure~\ref{net01}.

We first compute the total number of nodes $N_t$ and the total
number of edges $E_t$ in the tree $U_{t}$. Let $n_v(t)$ and $n_e(t)$
denote the numbers of nodes and edges created at step $t$,
respectively. Then, $n_v(0)=N_{0}=2$ and $n_e(0)=1$. By
construction, we have $n_v(t)=N_{t-1}$, thus
$N_{t}=n_v(t)+N_{t-1}=2\,N_{t-1}$. Considering the initial condition
$N_{0}=2$, we obtain $N_t=2^{t+1}$ and $n_v(t)=2^{t}$. Since $U_{t}$
is a tree, we have the following relation $E_t=N_t-1=2^{t+1}-1$. On
the other hand, at arbitrary step $t\geq 1$, the addition of each
new node leads to only new edge, thus $n_e(t)=n_v(t)=2^{t}$ for all
$t\geq 1$.

\section{Topological properties}
Topological features of a network are of fundamental significance to
understanding the complex dynamics taking place on it. Here we focus
on four important characteristics of the tree $U_{t}$, i.e., degree
distribution, average path length, betweenness distribution, and
degree correlations.

\subsection{Degree distribution}

The degree is the simplest and most intensively studied
characteristic of an individual node. The degree of a node $i$ is
the number of edges in the whole network connected to $i$. The
degree distribution $P(k)$ is defined as the probability that a
randomly selected node has exactly $k$ edges. Let $k_{i}(t)$ denote
the degree of node $i$ at step $t$. If node $i$ is added to the
network at step $t_i$, then by construction, $k_{i}( t_i)=1$. In
each of the subsequent time steps, a new node will be created
connected to $i$. Thus the degree $k_i(t)$ of node $i$ satisfies the
relation
\begin{equation}
k_{i}(t)=k_{i}(t-1)+1.
\end{equation}
Considering the initial condition $k_{i}( t_i)=1$, we obtain
\begin{equation}\label{Ki}
k_{i}(t)=t-t_{i}+1.
\end{equation}
Since the degree of each node has been obtained explicitly as in
Eq.~(\ref{Ki}), we can get the degree distribution via its
cumulative distribution~\cite{Ne03}
\begin{equation} \label{cumulative distribution1}
P_{cum}(k)=\sum_{k'=k}^{\infty}P(k'),
\end{equation}
which is the probability that the degree is greater than or equal to
$k$. An important advantage of the cumulative distribution is that
it can reduce the noise in the tail of probability distribution.
Moreover, for some networks whose degree distributions have
exponential tails: $P(\tilde{k}) \sim e^{-\tilde{k}/\kappa}$, the
cumulative distribution also has an exponential expression with the
same exponent:
\begin{equation} \label{cumulative distribution2}
P_{cum}(\tilde{k})=\sum_{k'=\tilde{k}}^{\infty}P(k')\sim
\sum_{k'=\tilde{k}}^{\infty}e^{-k'/\kappa}\sim
e^{-\tilde{k}/\kappa}.
\end{equation}
This makes exponential distributions particularly easy to detect
experimentally, by plotting the corresponding cumulative
distributions on semilogarithmic scales.

Using Eq.~(\ref{Ki}), we have
$P_{cum}(k)=\sum_{k'=k}^{\infty}P(k')\\= P\left
(t'\leq\tau=t+1-k\right)$. Hence
\begin{eqnarray}\label{cumulative distribution3}
P_{cum}(k)=\sum_{t'=0}^{\tau}\frac{n_v(t')}{N_{t}}
=\frac{2^{t+2-k}}{2^{t+1}}=2^{-(k-1)},
\end{eqnarray}
which decays exponentially with $k$. Thus the deterministic uniform
recursive tree is an exponential network, and has a similar form of
degree distribution as its stochastic version--- the random uniform
recursive tree~\cite{ZhZhZhGu08}.

\subsection{Average path length}

Average path length (APL) means the minimum number of edges
connecting a pair of nodes, averaged over all node pairs. It is
defined to be:
\begin{equation}\label{APL01}
  \bar{d}_t  = \frac{S_t}{N_t(N_t-1)/2}\,,
\end{equation}
where $S_t$ denotes the sum of the total distances between two nodes
over all pairs, that is
\begin{equation}\label{APL02}
  S_t = \sum_{i\neq j} d_{i,j}\,,
\end{equation}
where $d_{i,j}$ is the shortest distance between node $i$ and $j$.

Let $\Omega_{\text{new}}^{t}$ and $\Omega_{\text{old}}^{t}$
represent the set of nodes created at step $t$ or earlier,
respectively.  Then one can write the sum over all shortest paths
$S_{t}$ in network $U_t$ as
\begin{equation}\label{APL03}
  S_{t} = \sum_{\substack{i \in \Omega_{\text{new}}^{t},\,j\in
      \Omega_{\text{old}}^{t}}} d_{i,j}+\sum_{\substack{i \in \Omega_{\text{new}}^{t},\,j\in
      \Omega_{\text{new}}^{t}}} d_{i,j}+\sum_{\substack{i \in \Omega_{\text{old}}^{t},\,j\in
      \Omega_{\text{old}}^{t}}} d_{i,j},
\end{equation}
where the third term is exactly $S_{t-1}$, \emph{i.e.},
\begin{equation}\label{APL04}
  \sum_{\substack{i \in \Omega_{\text{old}}^{t},\,j\in
      \Omega_{\text{old}}^{t}}} d_{i,j}=S_{t-1}.
\end{equation}
By construction, we can obtain the following relations for the first
and second terms:
\begin{equation}\label{APL05}
  \sum_{\substack{i \in \Omega_{\text{new}}^{t},\,j\in
      \Omega_{\text{old}}^{t}}} d_{i,j}=N_{t-1}^{2}+2\,S_{t-1},
\end{equation}
\begin{equation}\label{APL06}
  \sum_{\substack{i \in \Omega_{\text{new}}^{t},\,j\in
      \Omega_{\text{new}}^{t}}} d_{i,j}=S_{t-1}+N_{t-1}(N_{t-1}-1).
\end{equation}
Substituting Eqs.~(\ref{APL04}),~(\ref{APL05}) and~(\ref{APL06})
into Eq.~(\ref{APL03}) and considering $N_{t}=2^{t+1}$, the total
distance is obtained to be
\begin{align}\label{APL07}
 S_t &= 4\,S_{t-1}+2\,N_{t-1}^{2}-N_{t-1}\nonumber \\
   &= 4^{t}\,S_0 + 2\,\sum_{i=0}^{t-1} 4^{t-1-i} N_i^{2}-\sum_{i=0}^{t-1} 4^{t-1-i} N_i\nonumber \\
&= 2t\cdot 4^{t}+2^{t}.
\end{align}
Inserting Eq.~(\ref{APL07}) into Eq.~(\ref{APL01}), we have
\begin{equation}\label{APL08}
  \bar{d}_t = \frac{2 (2t\cdot 4^{t}+2^{t})}{2^{t+1}(2^{t+1}-1)}=\frac{t\cdot
  2^{t+1}+1}{2^{t+1}-1}.
\end{equation}
In the infinite network size limit ($t \rightarrow \infty$),
\begin{equation}\label{APL09}
\bar{d}_{t} \cong \frac{\ln N_{t}}{\ln 2}-1,
\end{equation}
which means that the average path length shows a logarithmic scaling
with the size of the network, indicating a similar small-world
behavior as the URT~\cite{DoMeOl06} and the Watts-Strogatz (WS)
model~\cite{WaSt98}.

\subsection{Betweenness distribution}
 Betweenness of a node is the
accumulated fraction of the total number of shortest paths going
through the given node over all node pairs~\cite{Newman01,Ba04}.
More precisely, the betweenness of a node $i$ is
\begin{equation}
b_{i}=\sum_{j \ne i \neq k}\frac{\sigma_{jk}(i)}{\sigma_{jk}},
\end{equation}
where $\sigma_{jk}$ is the total number of shortest path between
node $j$ and $k$, and $\sigma_{jk}(i)$ is the number of shortest
path running through node $i$.

Since the considered network here is a tree, for each pair of nodes
there is a unique shortest path between
them~\cite{SzMiKe02,BoRi04,GhOhGoKaKi04,ZhZhChGuFaZh07}. Thus the
betweenness of a node is simply given by the number of distinct
shortest paths passing through the node. Then at time $t$, the
betweenness of a $\tau$-generation-old node $v$, which is created at
step $t-\tau+1$, denoted as $b_{t}(\tau)$ becomes
\begin{equation}\label{between01}
b_{t}(\tau)= \Theta_{t}^{\tau}\,\left[N_{t} -
\left(\Theta_{t}^{\tau}+1\right)\right]+\binom
{\Theta_{t}^{\tau}}{2}-\sum_{m=2}^{\tau-1} \binom
{\Theta_{t}^{m}+1}{2},
\end{equation}
where $\Theta_{t}^{\tau}$ denotes the total number of descendants of
node $v$ at time $t$, where the descendants of a node are its
children, its children¡¯s children, and so on. Note that the
descendants of node $v$ exclude $v$ itself. The first term in
Eq.~(\ref{between01}) counts shortest paths from descendants of $v$
to other vertices. The second term accounts for the shortest paths
between descendants of $v$. The third term describes the shortest
paths between descendants of $v$ that do not pass through $v$.

To find $b_{t}(\tau)$, it is necessary to explicitly determine the
descendants $\Theta_{t}^{\tau}$ of node $v$, which is related to
that of $v's$ children via~\cite{GhOhGoKaKi04,ZhZhChGuFaZh07}
\begin{equation}\label{child01}
\Theta_{t}^{\tau}= \sum_{j=1}^{\tau-1}\left(\Theta_{t}^{j}+1\right).
\end{equation}
Using $\Theta_{t}^{1}=0$, we can solve Eq.~(\ref{child01})
inductively,
\begin{equation}\label{child02}
\Theta_{t}^{\tau}= 2^{\tau-1}-1.
\end{equation}
Substituting the result of Eq.~(\ref{child02}) and $N_t=2^{t+1}$
into Eq.~(\ref{between01}), we have
\begin{equation}\label{between02}
b_{t}(\tau)=2^{t+\tau} -2^{t+1}-\frac{2}{3}\times2^{2(\tau -1)}+
\frac{2}{3},
\end{equation}
which is approximately equal to $2^{t+\tau}$ for large $\tau$. Then
the cumulative betweenness distribution is
\begin{eqnarray}\label{pcumb01}
P_{\rm cum}(b)&=&\sum_{\mu \leq t-\tau+1}\frac{n_v(\mu)}{N_t}
=\frac{2^{t+1}}{2^{t+\tau}} \approx {N_{t} \over b}\sim b^{-1},
\end{eqnarray}
which shows that the betweenness distribution exhibits a power law
behavior with exponent $\gamma_{b}=2$, the same scaling has been
also obtained for the URT~\cite{GoOhJeKaKi02} and the $m=1$ case of
the Barab\'asi-Albert (BA) model~\cite{BaAl99} describing a random
scale-free treelike network~\cite{SzMiKe02,BoRi04}. Therefore,
power-law betweenness distribution is not an exclusive property of
scale-free networks.

\subsection{Degree correlations}

An interesting quantity related to degree correlations~\cite{MsSn02}
is the average degree of the nearest neighbors for nodes with degree
$k$, denoted as $k_{\rm nn}(k)$ \cite{PaVaVe01,VapaVe02,ZhZh07}.
When $k_{\rm nn}(k)$ increases with $k$, it means that nodes have a
tendency to connect to nodes with a similar or larger degree. In
this case the network is defined as assortative
\cite{Newman02,Newman03c}. In contrast, if $k_{\rm nn}(k)$ is
decreasing with $k$, which implies that nodes of large degree are
likely to have near neighbors with small degree, then the network is
said to be disassortative. If correlations are absent, $k_{\rm
nn}(k)=const$.

For the deterministic uniform recursive tree, we can exactly
calculate $k_{\rm nn}(k)$. Except for the initial two nodes
generated at step 0, no nodes born at the same step, which have the
same degree, will be linked to each other. All links to nodes with
larger degree are made at the creation step, and then links to nodes
with smaller degree are made at each subsequent steps. This results
in the expression for $k=t+1-t_i$ ($t_i\geq 1$)
\begin{eqnarray}\label{knn01}
k_{\rm nn}(k)&=&{1\over n_v(t_i) k(t_i,t)} \Bigg[
  \sum_{t'_i=0}^{t'_i=t_i-1} n_v(t'_i)k(t'_i,t)\nonumber\\
  &\quad&+\sum_{t'_i=t_i+1}^{t'_i=t} n_v(t_i)
  k(t'_i,t)\Bigg],
\end{eqnarray}
where $k(t_i,t)$ represents the degree of a node at step $t$, which
was generated at step $t_i$. Here the first sum on the right-hand
side accounts for the links made to nodes with larger degree (i.e.\
$t'_i<t_i$) when the node was generated at $t_i$. The second sum
describes the links made to the current smallest degree nodes at
each step $t'_i>t_i$.

After some algebraic manipulations, Eq.~(\ref{knn01}) is simplified
to
\begin{align} \label{knn02}
k_{\rm nn}(k)=
\frac{t-t_i}{2}+\frac{t+3-t_i}{t+1-t_i}-\frac{2^{1-t_i}}{t+1-t_i}.
\end{align}
Writing Eq. (\ref{knn02}) in terms of $k$, it is straightforward to
obtain
\begin{align} \label{knn3}
k_{\rm nn}(k)= \frac{k+1}{2}+\frac{2}{k}-\frac{2^k}{k\cdot 2^t}.
\end{align}
Thus we have obtained the degree correlations for those nodes born
at $t_i\geq 1$. For the initial two nodes, each has a degree of
$k=t+1$, and it is easy to obtain
\begin{align}\label{knn4}
k_{\rm nn}(k=t+1)={1\over k}\, \sum_{t'_i=0}^{t'_i=t}
  k(t'_i,t)= \frac{k+1}{2}.
\end{align}
From Eqs.~(\ref{knn3}) and (\ref{knn4}), it is obvious that for
large network (i.e., $t\rightarrow \infty$), $k_{\rm nn}(k)$ is
approximately a linear function of $k$, which shows that the network
is assortative.

\section{Eigenvalues and eigenvectors of the Laplacian matrix}

As known from section 2, there are $2^{t+1}$ vertices in $U_t$. we
denote by $V_t$ the vertex set of $U_t$, i.e.,
$V_t=\{v_1,v_2,\ldots, v_{2^{t+1}}\}$. Let $\mathbf{A}_t=[a_{ij}]$
be the adjacency matrix of network $U_t$, where $a_{ij} =a_{ji}=1$
if nodes $i$ and $j$ are connected, $a_{ij} =a_{ji}=0$ otherwise,
then the degree of vertex $v_i$ is defined as  $d_{v_i}= \sum_{j\in
V_t}a_{ij} $. Let $\mathbf{D}_t =\text{diag} (d_{v_1},
d_{v_2},\ldots, d_{v_{2^{t+1}}})$ represent the diagonal degree
matrix of $U_t$, then the Laplacian matrix of $U_t$ is defined by
$\mathbf{L}_t=\mathbf{D}_t-\mathbf{A}_t$. For an arbitrary graph, it
is generally difficult to determine all eigenvalues and the
corresponding eigenvectors of it Laplacian matrix, but below we will
show that for $U_t$ one can settle this problem.

\subsection{eigenvalues}

We first study the Laplacian spectra of $U_t$ making use of an
iterative method~\cite{BaCoDaFi08}. By construction, it is easy to
find that the adjacency matrix $\textbf{A}_t$ and diagonal degree
matrix $\textbf{D}_t$ obeys the following relations:
\begin{equation}\label{matrix01}
\mathbf{A}_t=\left(\begin{array}{ccc}\textbf{A}_{t-1} &
\textbf{I}_{t-1}
\\\textbf{I}_{t-1} & {\textbf{0}}\end{array}\right)
\end{equation}
and
\begin{equation}\label{matrix02}
\mathbf{D}_t=\left(\begin{array}{ccc}\textbf{D}_{t-1}+\textbf{I}_{t-1}
& \textbf{0}
\\\textbf{0} & \textbf{I}_{t-1}\end{array}\right),
\end{equation}
 where each block is a $2^t\times 2^t$ matrix and
$\textbf{I}_{t-1}$ is identity matrix. Thus, according to the
definition of Laplacian matrix, we have the recursive relation
between $\textbf{L}_t$ and $\textbf{L}_{t-1}$ as
\begin{eqnarray}\label{matrix03}
\mathbf{L}_t=\mathbf{D}_t-\mathbf{A}_t &=&
\left(\begin{array}{ccc}\textbf{D}_{t-1}+\textbf{I}_{t-1}-\textbf{A}_{t-1}\quad
& \quad -\textbf{I}_{t-1}
\\-\textbf{I}_{t-1} & \quad\textbf{I}_{t-1}\end{array}\right)\nonumber\\
&=&\left(\begin{array}{ccc}\textbf{L}_{t-1}+\textbf{I}_{t-1} &
-\textbf{I}_{t-1}
\\-\textbf{I}_{t-1} & \textbf{I}_{t-1}\end{array}\right).
\end{eqnarray}
Then, the characteristic polynomial of $\textbf{L}_t$ is
\begin{eqnarray}\label{matrix04}
g_t(x)&=&\text{det}(x\textbf{I}_t-\textbf{L}_t)\nonumber\\
&=&\text{det}\left(\begin{array}{ccc}(x-1)\textbf{I}_{t-1}-\textbf{L}_{t-1}
& \textbf{I}_{t-1}
\\\textbf{I}_{t-1} & (x-1)\textbf{I}_{t-1}\end{array}\right)\nonumber\\
&=&\text{det}\left(\begin{array}{ccc}\left(x-1-\frac{1}{x-1}\right)\textbf{I}_{t-1}-\textbf{L}_{t-1}
& \textbf{I}_{t-1}
\\\textbf{0} & (x-1)\textbf{I}_{t-1}\end{array}\right),\nonumber\\
\end{eqnarray}
where the elementary column operations of matrix have been used.
According to the results in~\cite{Si00}, we have
\begin{eqnarray}\label{matrix05}
&\quad&g_t(x)\nonumber\\
&=&\text{det}\left(\left(x-1-\frac{1}{x-1}\right)\textbf{I}_{t-1}-\textbf{L}_{t-1}\right)\cdot \text{det}\big((x-1)\textbf{I}_{t-1}\big)\nonumber\\
&=&\big(x-1\big)^{2^t} \cdot
\text{det}\bigg(\left(x-1-\frac{1}{x-1}\right)\textbf{I}_{t-1}-\textbf{L}_{t-1}\bigg).
\end{eqnarray}
Thus, $g_t(x)$ can be written recursively as follows:
\begin{equation}\label{matrix06}
g_t(x)=\big(x-1\big)^{2^t}\cdot g_{t-1}\big(f(x)\big),
\end{equation}
where $f(x)=x-1-\frac{1}{x-1}$. This recursive relation given by
Eq.~\eqref{matrix06} is very important, from which we will determine
the complete Laplacian eigenvalues of $U_t$ and their corresponding
eigenvectors. Notice that $g_{t-1}(x)$ is a monic polynomial of
degree $2^t$, then the coefficient of $1/(x-1)^{2^t}$ in
$g_{t-1}(f(x))$ is 1, and hence 1 is the constant term of
$g_{t}(x)$. Consequently, 1 is never an eigenvalue of
$\textbf{L}_t$.

Note that $U_t$ has $2^{t+1}$ Laplacian eigenvalues, and all these
eigenvalues are distinct, which will be shown below. Let these
$2^{t+1}$ eigenvalues are $\lambda_1^t$, $\lambda_2^t$, \ldots,
$\lambda_{2^{t+1}}^t$, respectively.  For convenience, we presume
 $\lambda_1^t<\lambda_2^t< \ldots<\lambda_{2^{t+1}}^t$ and denote
by $\Delta_t$ the set of these eigenvalues of $U_t$, i.e.,
$\Delta_t$=\{$\lambda_1^t$, $\lambda_2^t$, \ldots,
$\lambda_{2^{t+1}}^t$\}.

From Eq.~\eqref{matrix06}, we have that for an arbitrary element in
$\Delta_{t-1}$, say $\lambda_{i}^{t-1} \in \Delta_{t-1}$, both
solutions of $x-1-\frac{1}{x-1}=\lambda_{i}^{t-1}$ are in
$\Delta_t$. In fact, equation $x-1-\frac{1}{x-1}=\lambda_{i}^{t-1}$
is equivalent to
\begin{equation}\label{matrix07}
x^2-(\lambda_{i}^{t-1}+2)\,x+\lambda_{i}^{t-1}=0.
\end{equation}
We use notations $\lambda_{i}^{t}$ and $\lambda_{i+2^t}^{t}$ to
represent the two solutions of Eq.~\eqref{matrix07}, since they
provide a natural increasing order of the Laplacian eigenvalues of
$U_t$, which can be seen from below argument. Solving this quadratic
equation, its roots are obtained to be
$\lambda_{i}^{t}=r_1(\lambda_{i}^{t-1})$ and
$\lambda_{i+2^t}^{t}=r_2(\lambda_{i}^{t-1})$, where the function
$r_1(\lambda)$ and $r_2(\lambda)$ satisfy
\begin{eqnarray}
r_1(\lambda)=\frac{1}{2}\left(\lambda+2-\sqrt{\lambda^2+4}\right),\label{matrix08}\\
r_2(\lambda)=\frac{1}{2}\left(\lambda+2+\sqrt{\lambda^2+4}\right).\label{matrix00}
\end{eqnarray}
Substituting each Laplacian eigenvalue of $U_{t-1}$ into Eqs.
\eqref{matrix08} and~\eqref{matrix00}, we can obtain the set
$\Delta_{t}$ of Laplacian eigenvalues of $U_{t}$. Since
$\Delta_{0}=\{0,2\}$, by recursively applying the functions provided
by Eqs. \eqref{matrix08} and~\eqref{matrix00}, the Laplacian
eigenvalues of $U_{t}$ can be determined completely.

It is obvious that both $r_1(\lambda)$ and $r_2(\lambda)$ are
monotonously increasing functions. On the other hand, since the
independent variables $\lambda$ here are greater than or equal to
zero, so $r_1(\lambda) \geq 0$ and $r_2(\lambda) > 0$. Furthermore,
$r_1(\lambda)=
(\lambda+2-\sqrt{\lambda^2+4})/2=(\lambda-\sqrt{\lambda^2+4})/2+1$,
where the term $(\lambda-\sqrt{\lambda^2+4})/2$ is clearly less than
zero, thus $r_1(\lambda)<1$. Similarly, we can show that
$r_2(\lambda) \geq 2$. Thus, for arbitrary fixed $\lambda'$,
$r_1(\lambda) < r_2(\lambda')$ holds for all $\lambda$. Then we have
the following conclusion: If the Laplacian eigenvalues set of
$U_{t-1}$ is $ \Delta _{t-1}=\{\lambda_1^{t-1}, \lambda_2^{t-1},
\ldots, \lambda_{2^t}^{t-1}\}$, then solving Eq.~\eqref{matrix07}
one can obtain the set of the Laplacian eigenvalues of $U_t$ to be
$\Delta _t=\{\lambda_1^t, \lambda_2^t, \ldots, \lambda_{2^t}^t,
\lambda_{2^t+1}^t, \ldots, \lambda_{2^{t+1}}^t \}$ where $0\leq
\lambda_1^t < \lambda_2^t < \ldots <\lambda_{2^t}^t<1<2 \leq
\lambda_{2^t+1}^t <\lambda_{2^t+2}^t < \ldots <
\lambda_{2^{t+1}}^t$. Therefore, all the $2^{t+1}$ Laplacian
eigenvalues of $U_t$ are different, which has never been previously
reported in other network models thus may have some far-reaching
consequences.

\subsection{eigenvectors}

Similarly to the eigenvalues, the eigenvectors of $\textbf{L}_t$
follow directly from those of $\textbf{L}_{t-1}$. Assume that
$\lambda$ is an arbitrary Laplacian eigenvalue of $U_t$, whose
corresponding eigenvectors is $\textbf{\emph{v}}$, then we can solve
equation ($\lambda\, \textbf{I}_t -\textbf{L}_t)\textbf{\emph{v}}=0$
to find the eigenvector $\textbf{\emph{v}}$. This equation can be
also written as
\begin{equation}\label{vector01}
(\lambda \textbf{I}_t
-\textbf{L}_t)\textbf{\emph{v}}=\left(\begin{array}{ccc}(\lambda-1)\textbf{I}_{t-1}-\textbf{L}_{t-1}
& \textbf{I}_{t-1}
\\\textbf{I}_{t-1} & (\lambda-1)\textbf{I}_{t-1}\end{array}\right)\left(\begin{array}{ccc}\textbf{\emph{v}}_1
\\\textbf{\emph{v}}_2\end{array}\right)=\textbf{0},
\end{equation}
where $\textbf{\emph{v}}_1$ and $\textbf{\emph{v}}_2$ are two
components of $\textbf{\emph{v}}$. Eq.~\eqref{vector01} leads to the
two following equations:
\begin{eqnarray}
\big((\lambda-1)\textbf{I}_{t-1}-\textbf{L}_{t-1}\big)\textbf{\emph{v}}_1+\textbf{\emph{v}}_2=\textbf{0},\label{vector02}
\\ \textbf{\emph{v}}_1+(\lambda-1)\textbf{\emph{v}}_2=\textbf{0}.\label{vector03}
\end{eqnarray}
Resolve Eq.~\eqref{vector03} to find
\begin{eqnarray}\label{vector04}
\textbf{\emph{v}}_2=-\frac{1}{\lambda-1}\textbf{\emph{v}}_1
\end{eqnarray}
Substituting Eq.~\eqref{vector04} into Eq.~\eqref{vector02} we have
\begin{eqnarray}\label{vector05}
\left[\left(\lambda-1-\frac{1}{\lambda-1}\right)\textbf{I}_{t-1}-\textbf{L}_{t-1}\right]\textbf{\emph{v}}_1=0,
\end{eqnarray}
which indicates that $\textbf{\emph{v}}_1$ is the solution of
Eq.~\eqref{vector02} while $\textbf{\emph{v}}_2$ is uniquely decided
by $\textbf{\emph{v}}_1$ via Eq.~\eqref{vector04}.

From Eq.~\eqref{matrix06} in preceding subsection, it is clear that
if $\lambda$ is an eigenvalue of Laplacian matrix $\textbf{L}_t$,
then $f(\lambda)=\lambda-1-\frac{1}{\lambda-1}$ must be one
eigenvalue of $\textbf{L}_{t-1}$. \big(Recall that if
$\lambda=\lambda_i^t \in \Delta_t$, then
$f(\lambda_{i}^t)=\lambda_{i}^{t-1}$ for $i\le2^t$, or
$f(\lambda_i^t)=\lambda_{i-2^t}^{t-1}$ for $i>2^t$\big). Thus,
Eq.~\eqref{vector05} together with Eq.~\eqref{matrix06} shows that
$\textbf{\emph{v}}_1$ is an eigenvector of matrix $\textbf{L}_{t-1}$
corresponding to the eigenvalue $\lambda-1-\frac{1}{\lambda-1}$
determined by $\lambda$, while
\begin{equation}
\textbf{\emph{v}}=\left(\begin{array}{ccc}\textbf{\emph{v}}_1
\\\textbf{\emph{v}}_2\end{array}\right)=\left(\begin{array}{ccc}\textbf{\emph{v}}_1\\-\frac{1}{\lambda-1}\textbf{\emph{v}}_1\end{array}\right)
\end{equation}
is an eigenvector of $\textbf{L}_t$ corresponding to the eigenvalue
$\lambda$.

Since for the initial graph $U_0$, its Laplacian matrix
$\textbf{L}_0$ has two eigenvalues 0 and 2 with respective
eigenvectors $(1,1)^\top$ and $(1,-1)^\top$.  By recursively
applying the above process, we can obtain all the eigenvectors of
the deterministic uniform recursive tree $U_t$.

\section{Conclusion and discussion}

We have investigated a deterministic model for the uniform recursive
tree, which is constructed in a recursive way. The model is actually
a deterministic variant of the intensively studied random uniform
recursive tree. We have presented an exhaustive analysis of many
properties of the considered model, and obtained the analytic
solutions for most of the topological features, including degree
distributions, average path length, betweenness distribution, and
degree correlations. Aside from their deterministic structures, the
obtained statistical characteristics are equivalent with its
corresponding random URT. Consequently, the DURT may provide useful
insight to the practices as URT.

Also, we have performed a detailed analysis of the complete
Laplacian spectra and eigenvectors of DURT by using the methods of
linear algebra and graph theory. We have fully characterized the
spectral properties and eigenvectors for DURT. We have shown that
all the eigenvalues and eigenvectors of the Laplacian for DURT can
be directly determined from those for the initial graph.
Interestingly, all the Laplacian eigenvalues have only one
multiplicity. To the best of knowledge, this property has never been
reported for other previously studied model. Finally, it should be
mentioned that although we have examined only a specific model, the
methods used here can be applicable to a larger type of networks.

\section*{Acknowledgment}

We thank Yichao Zhang for preparing this manuscript. This research
was supported by the National Basic Research Program of China under
grant No. 2007CB310806, the National Natural Science Foundation of
China under Grant Nos. 60496327, 60573183, 90612007, 60773123, and
60704044, the Shanghai Natural Science Foundation under Grant No.
06ZR14013, the China Postdoctoral Science Foundation funded project
under Grant No. 20060400162, the Program for New Century Excellent
Talents in University of China (NCET-06-0376), and the Huawei
Foundation of Science and Technology (YJCB2007031IN).


\end{document}